\documentclass[sigconf]{acmart}
\AtBeginDocument{%
  }

\usepackage{tikz}
\usepackage{algorithm}
\usepackage{algpseudocode}

\usepackage{booktabs}

\usepackage{listings}
\usepackage{xcolor} 

\usepackage{multirow}

\usepackage{subfigure}

\usepackage{url}

\usepackage{hyperref}

\begin{document}

\title{AutoVulnPHP: LLM-Powered Two-Stage PHP Vulnerability \\ Detection and Automated Localization}


\author{Zhiqiang Wang}
\authornote{Co-first authors.}       
\authornote{Corresponding authors.}  
\affiliation{
  \institution{Beijing Electronic Science and Technology Institute}
  \city{Beijing}
  \country{China}}
\email{wangzq@besti.eud.cn}

\author{Yizhong Ding}
\authornotemark[1]                   
\affiliation{
  \institution{Beijing Electronic Science and Technology Institute}
  \city{Beijing}
  \country{China}}
\email{chaoqunding5@gmail.com}

\author{Zilong Xiao}
\affiliation{
  \institution{Beijing Electronic Science and Technology Institute}
  \city{Beijing}
  \country{China}}
\email{602051399@qq.com}

\author{Jinyu Lu}
\affiliation{
  \institution{Beijing Electronic Science and Technology Institute}
  \city{Beijing}
  \country{China}}
\email{sdjzuljy2021@163.com}

\author{Yan Jia}
\affiliation{
  \institution{Nankai University}
  \city{Tianjin}
  \country{China}}
\email{jiay@nankai.edu.cn}

\author{Yanjun Li}
\affiliation{
  \institution{The 15th Research Institute of China Electronics Technology Group Corporation}
  \city{Beijing}
  \country{China}}
\email{liyanjun@itstec.org.cn}


\begin{abstract}
PHP's dominance in web development is undermined by security challenges: static analysis lacks semantic depth, causing high false positives; dynamic analysis is computationally expensive; and automated vulnerability localization suffers from coarse granularity and imprecise context. Additionally, the absence of large-scale PHP vulnerability datasets and fragmented toolchains hinder real-world deployment.

We present AutoVulnPHP, an end-to-end framework coupling two-stage vulnerability detection with fine-grained automated localization. SIFT-VulMiner (Structural Inference for Flaw Triage Vulnerability Miner) generates vulnerability hypotheses using AST structures enhanced with data flow. SAFE-VulMiner (Semantic Analysis for Flaw Evaluation Vulnerability Miner) verifies candidates through pretrained code encoder embeddings, eliminating false positives. ISAL (Incremental Sequence Analysis for Localization) pinpoints root causes via syntax-guided tracing, chain-of-thought LLM inference, and causal consistency checks to ensure precision.

We contribute PHPVD, the first large-scale PHP vulnerability dataset with 26,614 files (5.2M LOC) across seven vulnerability types. On public benchmarks and PHPVD, AutoVulnPHP achieves 99.7\% detection accuracy, 99.5\% F1 score, and 81.0\% localization rate. Deployed on real-world repositories, it discovered 429 previously unknown vulnerabilities, 351 assigned CVE identifiers, validating its practical effectiveness.

\end{abstract}





\maketitle

\section{Introduction}

Web applications constitute the backbone of modern digital infrastructure, facilitating services ranging from social media platforms to mission-critical business tools. Among server-side technologies, PHP retains a dominant position, powering over 76\% of known websites as of May 2024 \cite{w3techs2025}, including major ecosystems like WordPress \cite{wordpress2025} and Facebook \cite{facebook2024}. However, this ubiquity comes at a cost: PHP's dynamic typing and extensive legacy codebases create a fertile ground for security vulnerabilities. The 2024 OSSRA report \cite{synopsys2024} reveals that 96\% of open-source repositories contain vulnerable components, with 74\% harboring high-risk vulnerabilities. Consequently, the volume of web-related CVEs has surged by an annual average of 56\% over the past decade \cite{CVEProgram}, underscoring an urgent need for automated and precise defense mechanisms.

Despite this critical demand, developing an effective automated defense framework for PHP remains elusive due to four persistent challenges that existing solutions fail to address simultaneously:

The first obstacle is the absence of high-quality, large-scale PHP vulnerability datasets. Unlike languages such as Java and Python, which benefit from well-curated benchmarks, PHP lacks comprehensive, real-world vulnerability collections with ground-truth labels. This data scarcity severely hampers the development and validation of detection and localization approaches, forcing researchers to rely on limited or synthetic datasets that fail to capture the complexity of real-world PHP codebases.

The second major hurdle lies in the inherent tension between detection precision and computational efficiency. Conventional static analyzers, relying on pattern matching over Abstract Syntax Trees (AST) \cite{cheatham1964syntax} or Code Property Graphs (CPGs) \cite{yamaguchi2014modeling}, scale well but lack semantic depth. They frequently misinterpret benign constructs as malicious, resulting in prohibitive false-positive rates that overwhelm developers. Conversely, dynamic techniques like fuzzing and symbolic execution \cite{cadar2013symbolic} offer high precision but suffer from path explosion, making them computationally infeasible for large-scale, real-world codebases.

Compounding this issue is the semantic blindness prevalent in current detection methods. Unlike syntactic errors, PHP vulnerabilities often manifest as subtle logical flaws dependent on complex data flows and context \cite{duan2025oyster}. Traditional static methods and shallow machine learning models lack the reasoning capabilities to differentiate between sanitized and malicious data flows, leading to high false-negative rates when facing obfuscated or indirect vulnerability patterns \cite{xie2006static,wi2022hiddencpg,shi2024recurscan}. This challenge is particularly acute in PHP due to its permissive type system and complex taint propagation semantics.

Beyond detection, a significant gap persists regarding localization accuracy. Most existing toolchains are fragmented, detection tools rarely integrate seamlessly with localization tools \cite{hu2025sok}. While Large Language Models (LLMs) have shown potential in code generation \cite{li2025tuni,cheng2025gibberish,cheng2024gibberish}, applying them directly to vulnerability localization is risky. Without strict constraints, LLMs suffer from “hallucinations” \cite{cheng2025speaker,cheng2025usmid,cheng2024unimodal}, frequently providing localization results that are syntactically plausible but fail to pinpoint the root cause of vulnerabilities or, worse, introduce new vulnerabilities \cite{wei2022chain}. The integration of detection confidence scores with localization strategies remains largely unexplored in prior work.

To overcome these distinct limitations, we propose AutoVulnPHP, a comprehensive framework that integrates a cascaded two-stage detection pipeline with a constraint-aware automated localization mechanism, grounded in a newly constructed large-scale dataset. Our approach is methodically designed to dismantle the barriers of data scarcity, computational inefficiency, and semantic ambiguity through the following specific interventions:

First, to resolve the tension between computational efficiency and precision, we introduce a coarse-to-fine cascaded detection strategy. The first stage, SIFT-VulMiner, addresses the computational bottleneck by employing a lightweight, structure-centric analysis. By leveraging AST-derived features enhanced with data flow, it rapidly filters the search space, acting as a high-recall sieve that discards benign code while retaining potential threats. This is immediately followed by SAFE-VulMiner, a semantic-focused second stage that applies computationally intensive pre-trained code encoders (CodeBERT) and risk-biased attention mechanisms only to the narrowed candidates. This design effectively bridges the gap, achieving the scalability of static analysis while approaching the precision of dynamic methods. Second, to tackle the issue of inaccurate localization, we introduce the ISAL framework. Unlike unconstrained generative models, ISAL treats vulnerability localization as a constrained optimization problem. It combines deterministic micro-templates with the reasoning capabilities of Chain-of-Thought (CoT) LLMs, enforcing strict semantic and security constraints to ensure that generated patches are not only syntactically correct but also functionally secure. Finally, underpinning these modules is PHPVD, our contribution of the first large-scale, high-quality PHP vulnerability dataset, which provides the necessary ground truth to train these deep learning components effectively.

Existing methodologies fail to achieve this equilibrium due to inherent structural and probabilistic limitations. Traditional static analysis and Graph Neural Network (GNN) based approaches, such as HiddenCPG \cite{wi2022hiddencpg}, rely heavily on subgraph isomorphism and rigid code property graphs. While scalable, they suffer from a “structural rigidity” problem: they struggle to detect vulnerabilities that manifest as subtle logical flaws within structurally valid code, or they fail when code patterns deviate slightly from predefined graph signatures due to obfuscation or coding style variations. Similarly, approaches like RecurScan \cite{shi2024recurscan}, while effective for recurring patterns, lack the generalization capability to identify novel or context-dependent zero-day vulnerabilities. On the other end of the spectrum, pure Large Language Model based approaches often treat vulnerability localization as a free-form text generation task. Without the guidance of symbolic constraints or intermediate representations, these models are prone to the “stochasticity trap”, where the probabilistic nature of token generation leads to plausible-looking but functionally erroneous “hallucinations”, introducing new bugs or failing to resolve the root cause.

AutoVulnPHP distinguishes itself by explicitly coupling structural priors with semantic reasoning. Our core innovation lies in the recognition that vulnerability detection is not a single-step classification task but a progressive refinement process. By decoupling structural hypothesis generation (Stage I) from semantic verification (Stage II), we ensure that heavy semantic reasoning is allocated efficiently, avoiding the resource waste seen in monolithic deep learning detectors. Furthermore, our localization module, ISAL, innovates by inverting the typical LLM workflow: instead of letting the LLM dictate the localization outcome, we use the LLM to fill in context-specific details within a template-guarded, constraint-driven framework. This human-in-the-loop inspired design—automated via iterative refinement—ensures that the creativity of LLMs is confined within the boundaries of security best practices, fundamentally solving the reliability issues that plague current automated vulnerability localization tools.


To validate the practical utility of our framework, we deployed AutoVulnPHP in a real-world mining workflow. By deploying our pipeline on massive open-source PHP repositories, we achieved substantial real-world impact: AutoVulnPHP discovered 429 previously unknown vulnerabilities, 351 of which have already been assigned CVE identifiers. These results, combined with a 99.7\% detection accuracy on public benchmarks, definitively demonstrate the framework's effectiveness and practical applicability.

\textbf{Contributions.} This paper makes the following contributions:
\begin{itemize}
    \item \textbf{Two-stage vulnerability detection.} We introduce a cascaded detector with structure-first hypothesis generation and semantics-first confirmation, coupled by calibrated score fusion. The first stage, SIFT-VulMiner, parses code into AST-derived and flow-aware slices and uses LLM-guided structural scoring to propose vulnerability candidates. The second stage, SAFE-VulMiner, performs semantic verification with a pretrained code encoder to filter noise and refine the candidates. This design prunes noise, preserves recall, and reduces false positives, yielding 99.7\% detection accuracy and a 99.5\% F1 score on public benchmarks.
    \item \textbf{ISAL framework.} We introduce ISAL, an incremental sequence analysis and localization system that mixes templates with LLMs, enforces semantic and security constraints, and achieves an 81\% localization rate in real-world projects.
    \item \textbf{Large-scale PHP vulnerability dataset.} We construct PHPVD, the first comprehensive dataset of 26,614 PHP files containing 5.2M lines of code across seven vulnerability types, addressing the critical data scarcity in PHP security research. Using AutoVulnPHP, we discovered 429 previously unknown vulnerabilities, of which 351 were assigned CVE identifiers.
\end{itemize}

\section{Related Work}

\subsection{PHP Vulnerability Discovery}

Vulnerability detection research has historically diverged based on the underlying paradigms of programming languages, with distinct approaches tailored to C/C++, Java, and Python. In system languages like C and C++, the primary security focus lies in memory safety, where researchers prioritize detecting buffer overflows and pointer corruption through symbolic execution and fuzzing \cite{cadar2013symbolic,chipounov2012s2e}. Conversely, strongly typed languages like Java benefit from explicit type definitions and compiled structures, enabling static analyzers to construct precise control flow graphs and enforce type-state properties effectively \cite{saccente2019project}. However, PHP presents a unique and more elusive set of challenges that render these generic approaches insufficient. As a dynamically typed, interpreted scripting language, PHP relies heavily on implicit type juggling, “variable variables,” and runtime evaluation (e.g., eval, include), which obscure the data flow until execution time. Xie and Aiken \cite{xie2006static} identified early on that this dynamic nature creates a disconnect between static representation and runtime behavior, causing traditional static analysis tools—designed for rigid type systems—to suffer from path explosion and semantic blindness when applied to PHP. Unlike Python, which enforces strong dynamic typing, PHP’s loose comparison rules introduce specific logical flaws (e.g., magic hash collisions) that are rarely seen in other ecosystems. Consequently, generic multilingual detectors often fail to capture the semantic nuances of PHP web applications, necessitating a specialized framework like AutoVulnPHP that explicitly models these language-specific dynamic features.

\subsection{Static Analysis for Vulnerability Detection}
Static analysis examines source code without execution and searches for syntactic and data-flow patterns that indicate vulnerabilities. Among static analysis techniques, AST-based methods parse PHP into structured representations. This enables precise pattern matching and control- and data-flow checks. Feng et al.\ \cite{feng2020efficient} extract AST features and train bidirectional GRUs \cite{cho2014learning} to detect CWE-type vulnerabilities across 118 categories with high accuracy, without relying on rule-based engines. These structural representations preserve program-logic hierarchies and work well for defects tied to improper structure or flawed control paths.

Code property graphs (CPG) combine AST, control flow, and data flow into a unified graph and enable scalable detectors, for example the CPG-based system by Wi \cite{wi2022hiddencpg} that locates diverse web flaws in billion-node graphs. Tools such as RIPS leverage PHP tokenization and interprocedural taint analysis to identify SQL injection, XSS, and 13 other common flaw classes in large codebases. This multidimensional integration allows CPG-based approaches to track explicit code relationships and implicit data dependencies that are critical to vulnerability manifestation.

Recent work adopts deep learning \cite{bengio2021deep,li2024single} with static analysis \cite{jiang2025never,jia2025omnisafebench}. Saccente et al.\ \cite{saccente2019project} employ LSTM-RNNs over AST-derived sequences and achieve over 90\% detection accuracy on 24 of 29 CWE families. Sequence modeling \cite{vaswani2017attention} enables automatic discovery of complex vulnerability patterns that would otherwise require extensive manual feature engineering in traditional static analysis frameworks.

However, static analysis often struggles with semantically subtle bugs, exhibits high false-positive rates, and requires extensive manual rule engineering. As a result, while static analysis is scalable and precise for structural patterns, it is insufficient for the detection of deep semantic vulnerabilities.

\begin{figure*}[!t]
\centering
\includegraphics[width=0.8\textwidth]{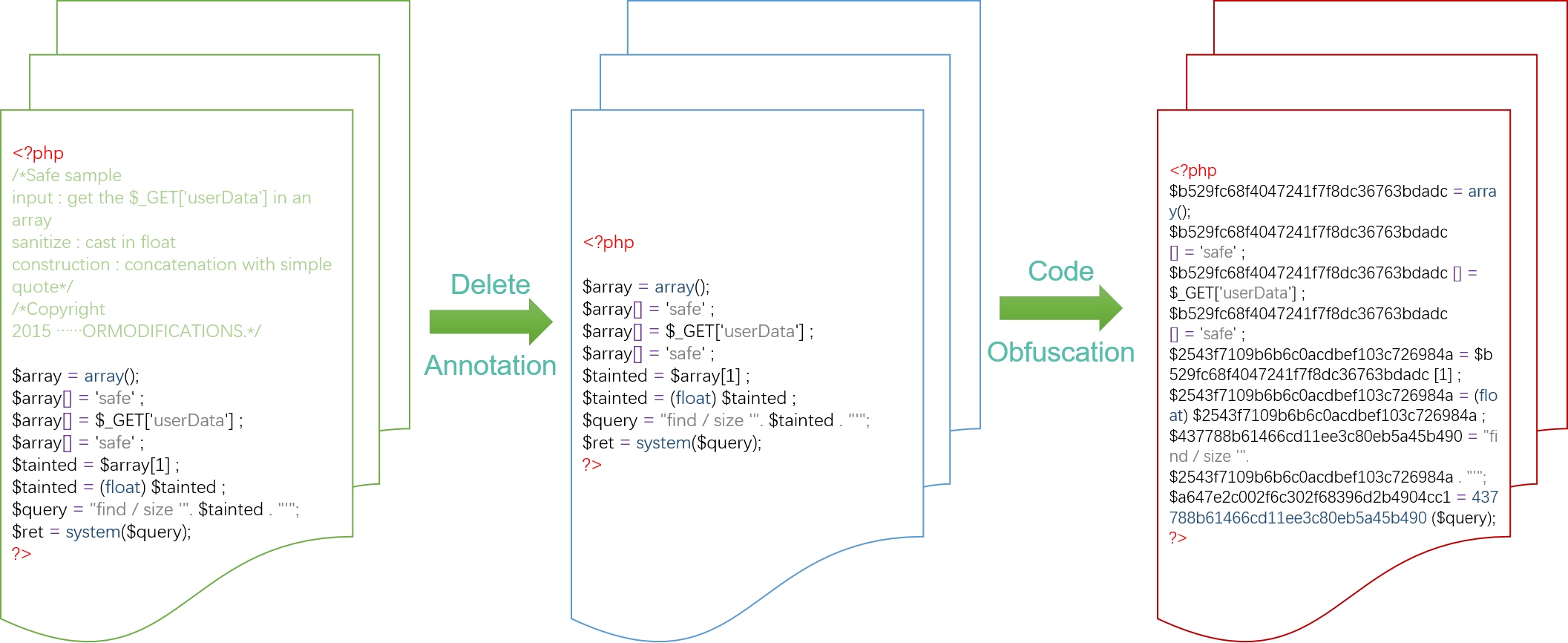}
\caption{Illustration of the Data Augmentation Process. The workflow transforms original PHP samples through annotation removal and code obfuscation to enhance model robustness.}
\label{code_process}
\end{figure*}

To further optimize coverage guidance, Markov chain guided fuzzing \cite{bohme2016coverage} models execution paths as probabilistic transitions and prioritizes mutations likely to trigger new paths, increasing path diversity and revealing subtle vulnerabilities in real services. This approach excels when input structures are complex because it adapts to runtime behaviors rather than relying on static heuristics.

For vulnerabilities requiring deeper semantic reasoning, hybrid symbolic execution frameworks, such as MAYHEM \cite{cha2012unleashing}, integrate symbolic constraint solving with concrete execution. By combining symbolic variables with real-time memory modeling, they can generate exploit payloads for zero-day vulnerabilities while remaining scalable to large applications. Similarly, S2E embeds symbolic execution in a virtual machine to explore user-space and kernel-space paths through dynamic binary translation \cite{chipounov2012s2e}.

For automated testing at scale, tools such as FuzzGen generate input harnesses by analyzing API dependencies and parameter constraints, achieving higher branch coverage and discovering new vulnerabilities more efficiently than manually written fuzzers \cite{ispoglou2020fuzzgen,wu2025sokunderstandingnewsecurity,cheng2025inverse}. These advances highlight a shift toward adaptive, feedback-driven dynamic testing that balances exploration depth with computational efficiency. Even with these advances, efficiency bottlenecks remain at scale. Thus, while dynamic testing improves coverage, it remains limited in scalability and cannot provide an end-to-end solution on its own.

\subsection{Machine Learning Based Approaches}
Traditional machine-learning methods \cite{rabheru2022hybrid,li2024identity} rely on structured feature engineering and shallow models for vulnerability detection. Existing approaches often convert AST into structured representations to extract hierarchical features using sequence models like GRUs. This structural encoding preserves code context while reducing raw-sequence complexity, enabling better capture of syntactic vulnerability patterns.

These methods emphasize explicit modeling of code structure and improve generalization through data augmentation \cite{ge2025mrfd,ge2025framemind,yang2024lever} such as variable renaming and syntax-tree transformation. AST is generated with tools such as PHP-Parser, which provide language-specific syntax coverage and error-resilient parsing for incomplete snippets. In addition, attention-based models can focus on high-risk code regions, but feature extraction still depends on handcrafted rules \cite{risse2024uncovering}. For instance, some frameworks employ AST-subtree matching to identify patterns such as SQL injection \cite{siddiq2021sqlifix}, which requires domain-specific rules for node traversal. 

Despite these advances, ML approaches \cite{wei2025dual,cheng2024rlrf} remain dependent on manual feature design, suffer from data imbalance, and often miss fine-grained semantic context. Their reliance on handcrafted features limits robustness against previously unseen vulnerability patterns.

\subsection{Large Language Model Based Approaches}
Pretrained code models (e.g., CodeBERT \cite{feng2020codebert} and CodeT5 \cite{wang2021codet5}) learn rich semantics from large-scale code corpora \cite{chengpbi}. Existing approaches leverage these models to extract semantic features and apply Transformer-based classifiers for end-to-end vulnerability detection. By leveraging pretrained knowledge of programming languages, these models encode syntax-aware and context-sensitive patterns without explicit rules \cite{cao2024agr,cheng2024deceiving}, an advantage observed across multiple real-world vulnerability benchmarks \cite{zhao2025strata, cheng2025pbi, pei2025selfprompt, teng2024heuristic, jia2025omnisafebench}.

These methods avoid manual feature engineering but rely on preprocessing—such as removing comments, normalizing strings and numbers, and renaming variables—to reduce noise. In particular, pipelines that remove comments, replace numeric constants and string literals with canonical identifiers, and randomize identifiers while preserving high-risk tokens such as \texttt{eval} and \texttt{system} improve the signal-to-noise ratio before modeling \cite{kalouptsoglou2024vulnerability}. LLMs also perform well in vulnerability localization when combined with reasoning mechanisms \cite{cao2025agr,cheng2025ecoalign}. For example, VRpilot \cite{kulsum2024case} uses CoT prompting and iteratively refines patches with feedback from compilers or sanitizers, increasing the rate of semantically correct fixes. However, LLM-based repair still faces limitations \cite{wang2025introspective, cheng2025talk,li2025clipscore,cheng2025hair}, including restricted context windows for long code, limited interpretability of fixes, and inconsistent patch quality. Many methods work only in narrow settings, so scale and reliability remain open challenges.

\begin{figure*}[!t]
\centering
\includegraphics[width=0.8\textwidth]{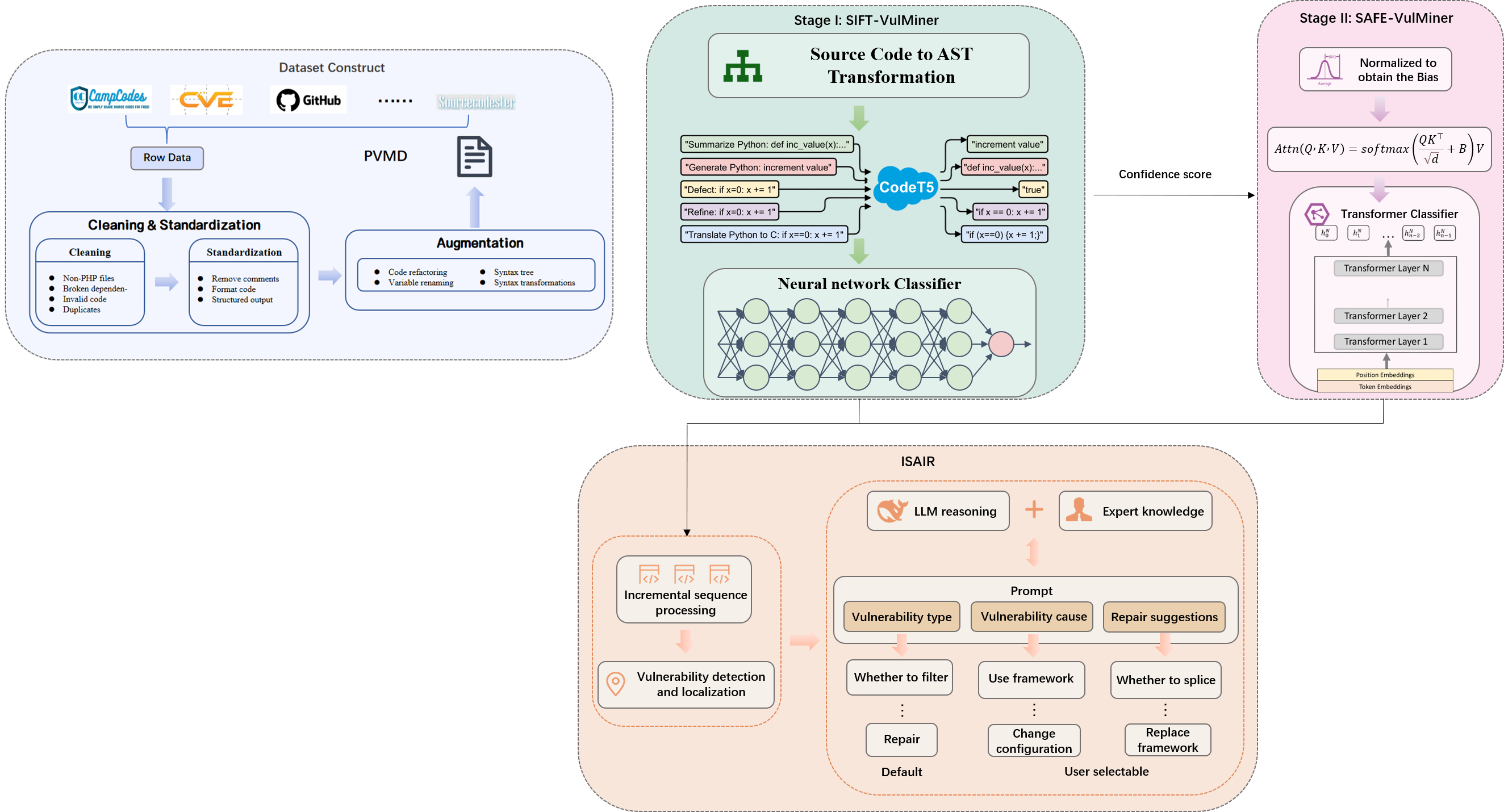}
\caption{Overview of the proposed framework.}
\label{pipeline}
\end{figure*}

\section{Methodology}

AutoVulnPHP establishes an integrated framework for detecting and localizing vulnerabilities in PHP applications. As visualized in Figure~\ref{pipeline}, our approach combines a systematically constructed dataset (PHPVD) with a two-stage cascaded detection pipeline and a constraint-aware automated localization module. The detection phase first generates vulnerability hypotheses through structural analysis, then verifies them via semantic validation. For confirmed vulnerabilities, the localization component generates context-aware localization results through template-guided generation, LLM inference, and constraint-driven validation.

\subsection{Problem Formulation}

Let $C$ denote a PHP codebase and $s \subseteq C$ a code slice. We formulate vulnerability detection as a binary classification task: predict $y \in \{0, 1\}$ indicating whether $s$ is vulnerable. Formally, we learn a detector $f: C \to \{0, 1\}$ that maps code to vulnerability labels. For localization, given a vulnerable slice $s$ with context $C(s)$, we seek an optimal patch $\hat{s}$ that eliminates vulnerability while preserving semantics, formulated as a constrained optimization problem detailed in Section~\ref{subsec:isair}.

\subsection{Dataset Construction}
\label{subsec:dataset_construction}

\begin{figure*}[!t]
\centering
\includegraphics[width=0.8\textwidth]{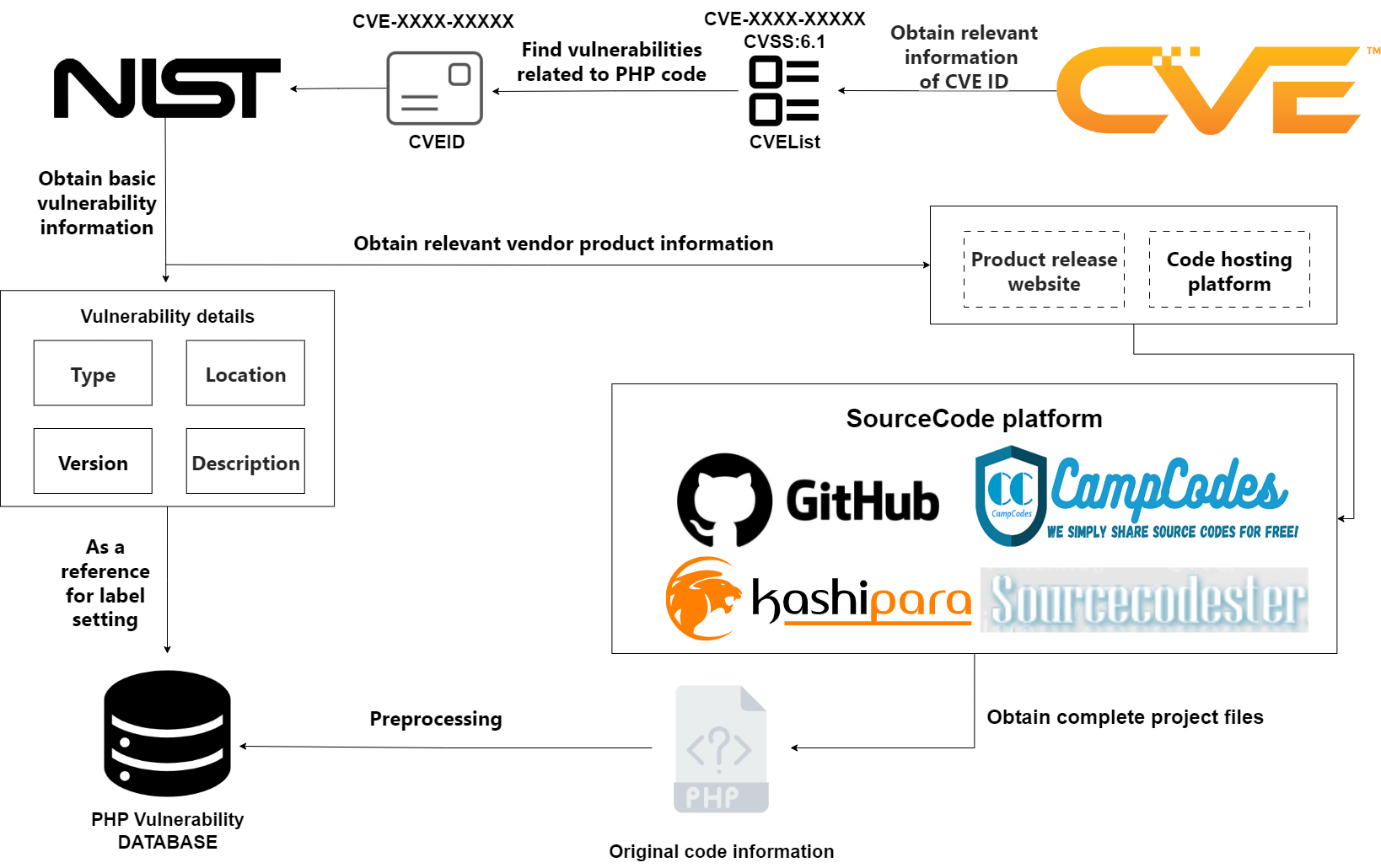}
\caption{Flowchart of Vulnerability Mining Dataset Collection.}
\label{fig_data_collection}
\end{figure*}

To address limitations in existing PHP vulnerability datasets—such as limited scale, missing real-world dependencies, and class imbalance—we constructed the PHP Vulnerability Dataset (PHPVD), a comprehensive and high-quality resource. The construction process emphasizes real-world relevance, diversity, and scalability. Its workflow, shown in Figure~\ref{fig_data_collection}, supports robust training and evaluation of deep learning models for vulnerability detection.

\textbf{Data Sources and Collection Process.} 
We first collected PHP-related CVE entries from the National Vulnerability Database maintained by the National Institute of Standards and Technology for the years 2020 to 2025. Each entry includes affected vendors, impacted versions, CWE categories, and disclosure dates. Using this information, we locate the corresponding open source projects on GitHub, Sourcecodester, Kashipara, and CampCodes. This multi-source collection strategy ensures broad coverage across diverse application domains and development styles. We also added representative cases from the Open Worldwide Application Security Project to include well-documented attack patterns and remediation practices.

\textbf{Data Cleaning and Standardization.} 
AutoVulnPHP establishes an integrated framework for detecting and localizing vulnerabilities in PHP applications. As visualized in Figure~\ref{pipeline}, our approach begins with PHPVD, a systematically constructed dataset that addresses data scarcity through rigorous collection, cleaning, and augmentation strategies. To detect vulnerabilities, we adopt a two-stage cascaded architecture. First, SIFT-VulMiner analyzes structural properties by constructing AST enriched with control and data flow information, extracting path-level and context-aware structural features, and applying a recall-oriented classifier to propose high-coverage vulnerability hypotheses. Second, SAFE-VulMiner revisits these candidates using semantic embeddings, risk-biased attention over critical code regions, and a precision-oriented scoring head, thereby pruning spurious alarms while preserving genuinely risky cases. For confirmed vulnerabilities, the ISAL localization module generates context-aware localization results through template-guided edits, LLM inference, and constraint-driven validation, utilizing an iterative refinement process that analyzes failure signals to progressively enhance localization accuracy and robustness.

\textbf{Data Augmentation.}
To address the severe class imbalance where vulnerable files are less than 1\% of the dataset, we apply a multifaceted augmentation strategy that focuses on vulnerable samples. This approach not only increases the number of vulnerable instances but also enhances the structural and semantic diversity of the training data, thereby improving the model's generalization capability. We employ three primary augmentation techniques: (1) \textit{Code Refactoring}, which involves transforming loop constructs into equivalent recursive implementations and reordering conditional branches without altering the program's functional behavior; (2) \textit{Variable Renaming}, where meaningful variable identifiers (e.g., \texttt{username}) are systematically replaced with syntactically diverse but semantically neutral alternatives (e.g., \texttt{user\_name}, \texttt{\$u\_name}, or randomly generated names), ensuring the model learns to identify vulnerabilities independent of specific naming conventions; and (3) \textit{Syntax Tree Transformation}, which applies structural modifications to the AST by reordering function calls, simplifying complex expressions, or altering control flow patterns, thereby generating syntactically distinct yet functionally equivalent variants of the original vulnerable code. These transformations are carefully designed to preserve the core vulnerability pattern while introducing significant syntactic variation, forcing the model to learn deeper, more robust representations of the underlying security flaws.

\textbf{Dataset Statistics.}
The final PHPVD comprises 26,614 PHP files, totaling over 5.2 million lines of code, covering seven common vulnerability types: IDOR (Insecure Direct Object Reference), Injection, SDE (Sensitive Data Exposure), SM (Security Misconfiguration), URF (Unvalidated Redirects and Forwards), XSS (Cross-Site Scripting), and File Inclusion. All files are labeled according to whether they are associated with a CVE. Specifically, files associated with a CVE are labeled as vulnerable (positive), while others are considered secure (negative).

\subsection{Stage I: Structural Inference for Flaw Triage Vulnerability Miner}
\label{subsec:lat-vulminer}

SIFT-VulMiner serves as a coarse-grained, high-recall filter designed to rapidly prune the search space. Its primary objective is to identify vulnerability hypotheses based on code structure and data flow dependencies, minimizing the risk of missing potential threats (False Negatives). The workflow begins by parsing PHP source code into a flow-enhanced graph representation, which is then linearized and mapped into a vector space by a pre-trained encoder. Finally, a sequence-based classifier assesses the structural likelihood of vulnerability, flagging suspicious candidates for the subsequent stage.

\textbf{Flow-Enhanced AST Construction.}
Standard Abstract Syntax Trees (ASTs) capture grammatical hierarchy but often obscure the propagation of tainted data. To address this, we parse the raw PHP code using \textit{PHP-Parser} to generate a normalized AST and augment it with two types of semantic edges: \textit{Control Flow Edges}, which connect nodes representing sequential execution, branches, and loops; and \textit{Data Flow Edges}, which link variable definitions to their usages. This augmented graph structure enables the model to explicitly trace how potentially malicious inputs traverse through program logic.

\textbf{Linearized Structural Embedding.}
To leverage the power of pre-trained language models on graph data, we employ a depth-first traversal strategy to linearize the augmented AST into a token sequence. This sequence preserves the structural integrity of the code while adhering to the input format of our encoder, \textbf{CodeT5}. By feeding the linearized sequence into CodeT5, we obtain a dense, continuous vector representation (embedding) that encapsulates both the syntactic structure and the long-range dependencies captured by the flow enhancements.

\textbf{Recall-Prioritized Hypothesis Generation.}
The extracted embeddings are fed into a Gated Recurrent Unit (GRU) classifier. Unlike traditional classifiers that aim for balanced accuracy, this module is optimized with a recall-oriented loss function. It deliberately lowers the decision threshold to accept a higher rate of false positives, ensuring that even subtle or obfuscated vulnerability patterns are retained as hypotheses for the fine-grained analysis in Stage II.

\begin{algorithm}[t]
\caption{Stage I: Structural Hypothesis Generation}
\label{alg:latvulminer}
\begin{algorithmic}[1]
\Require Source code corpus $\mathcal{C}$; AST converter $\mathsf{Conv}$; LLM encoder $\mathsf{Enc}$; classifier $\mathsf{Clf}$
\Ensure Vulnerability candidates with scores
\For{each file $c \in \mathcal{C}$}
    \State $\mathsf{AST}_c \gets \mathsf{Conv}(c)$ 
    \State $x_c \gets \text{Augment\_Flows}(\mathsf{AST}_c)$ 
    \State $\mathbf{z}_c \gets \mathsf{Enc}(\text{Linearize}(x_c))$ 
    \State $\text{score}_c \gets \mathsf{Clf}(\mathbf{z}_c)$ 
\EndFor
\State \Return $\{c, \text{score}_c : \text{score}_c > \tau_1\}$
\end{algorithmic}
\end{algorithm}

\subsection{Stage II: Semantic Analysis for Flaw Evaluation Vulnerability Miner}
\label{subsec:cbt-vulminer}

The second stage, SAFE-VulMiner, acts as a fine-grained verifier dedicated to eliminating false positives (High Precision). It conducts a deep semantic analysis on the candidates proposed by Stage I, utilizing a risk-aware attention mechanism to focus on security-critical operations. This stage creates a focused evaluation pipeline: it first cleanses the input to remove irrelevant noise, then applies a specialized Transformer model that heavily weights risky code regions (such as SQL execution or system calls) to make a definitive vulnerability judgment.

\textbf{Context Normalization and Noise Filtering.}
To prevent the model from overfitting to irrelevant stylistic features, we perform rigorous preprocessing on the candidate code slices. This involves removing non-executable artifacts such as comments and blank lines, as well as standardizing user-defined identifiers (e.g., variable names and function aliases). This normalization ensures that the semantic analysis focuses purely on the code's logic and behavior rather than its formatting.

\textbf{Risk-Biased Attention Mechanism.}
Standard self-attention mechanisms treat all code tokens with equal distinctiveness, often diluting the signal of vulnerable statements in large codebases. We introduce a \textit{Risk-Biased Self-Attention} layer that explicitly integrates prior security knowledge. Let $Q, K, V \in \mathbb{R}^{n \times d}$ be the query, key, and value matrices, and let $\mathbf{B} \in \mathbb{R}^{n \times n}$ be a pre-computed risk matrix where high values correspond to known sensitive tokens (e.g., \texttt{eval}, \texttt{query}, \texttt{\$\_GET}). The attention is computed as:
\[
\mathrm{Attn}(Q,K,V)=\mathrm{softmax}\!\left(\frac{QK^{\top}}{\sqrt{d}}+\mathbf{B}\right)V
\]
The bias term $\mathbf{B}$ forces the attention heads to pay significantly more attention to high-risk operations, effectively suppressing the influence of benign code context.

\textbf{Precision-Driven Semantic Verification.}
The output of the risk-biased attention layer is processed by a Transformer-based classifier trained with a strong penalty for false alarms. By leveraging the semantic understanding of \textbf{CodeBERT} combined with our risk-biased attention, this module captures high-order interactions between input sources and sensitive sinks. It outputs a final, high-confidence probability score, confirming whether the structural hypothesis from Stage I is indeed a genuine vulnerability.

\begin{algorithm}[t]
\caption{Stage II: Semantic Verification and Refinement}
\label{alg:cbt}
\begin{algorithmic}[1]
\Require Candidates $\mathcal{H}$ from Stage I; cleaner $\mathsf{Clean}$; encoder $\mathsf{Enc}$; Transformer $\mathsf{Trans}$; risk bias $\mathbf{B}$
\Ensure Verified vulnerabilities with refined scores
\State $\mathcal{D}' \gets \mathsf{Clean}(\mathcal{H})$ 
\For{each candidate $h \in \mathcal{D}'$}
    \State $t_h \gets \text{Tokenize}(h)$
    \State $\mathbf{e}_h \gets \mathsf{Enc}(t_h)$ 
    \State $\mathbf{A} \gets \mathrm{softmax}\!\left(\frac{\mathbf{e}_h\mathbf{e}_h^{\top}}{\sqrt{d}}+\mathbf{B}\right)$ 
    \State $\text{score}_h^{\text{SAFE}} \gets \mathsf{Trans}(\mathbf{A}\mathbf{e}_h)$ 
\EndFor
\State \Return verified vulnerabilities with refined scores
\end{algorithmic}
\end{algorithm}

\subsection{Two-Stage Cascaded Workflow}

The two stages operate sequentially: Stage I generates high-recall structural hypotheses; Stage II confirms high-precision semantic vulnerabilities. Their outputs are combined via weighted fusion:
\[
\text{score}^{\text{final}} = \lambda \cdot \text{score}^{\text{I}} + (1 - \lambda) \cdot \text{score}^{\text{II}},
\]
where $\lambda \in [0,1]$ is tuned on validation data. Stage I restricts Stage II's input space, reduces computation on low-risk regions, and guides attention toward structurally risky positions, while Stage II provides semantic confirmation that stabilizes predictions when Stage I is uncertain.

\begin{algorithm}[t]
\caption{Two-Stage Cascaded Detection Pipeline}
\label{alg:two_stage_pipeline}
\begin{algorithmic}[1]
\Require Source code corpus $C$; Stage I detector $f_I$; Stage II detector $f_{II}$; threshold $\tau$
\Ensure Final vulnerability scores
\State $\mathcal{H}_{\text{I}} \gets f_I(C)$ 
\For{each slice $s_i \in \mathcal{H}_{\text{I}}$ with high score}
    \State $\text{score}_i^{\text{II}} \gets f_{II}(s_i)$ 
    \State $\text{score}_i^{\text{final}} \gets \lambda \cdot \text{score}_i^{\text{I}} + (1-\lambda)\cdot \text{score}_i^{\text{II}}$
\EndFor
\State \Return vulnerabilities with $\text{score}^{\text{final}} > \tau$
\end{algorithmic}
\end{algorithm}

\subsection{Incremental Sequence Analysis for Localization}
\label{subsec:isair}

Given a detected vulnerability, we formalize automated localization as a constrained optimization problem. Given vulnerable slice $s$ and context $C(s)$, produce a precise localization result $\hat{s}$ that pinpoints the vulnerability location, preserves semantics, and minimizes manual effort:

\[
\hat{s} \;=\; \arg\min_{s' \in \mathcal{G}(s,C(s))}\; \mathcal{L}_{\mathrm{safe}}(s') + \lambda_1\,\mathcal{L}_{\mathrm{func}}(s',C(s)) + \lambda_2\,\mathcal{L}_{\mathrm{edit}}(s',s),
\]

where $\mathcal{G}(s,C(s))$ is the candidate space, $\mathcal{L}_{\mathrm{safe}}$ measures residual risk, $\mathcal{L}_{\mathrm{func}}$ penalizes semantic deviation, and $\mathcal{L}_{\mathrm{edit}}$ captures edit distance.

\textbf{Constraint Extraction.}
ISAL derives constraints from intermediate representation (control/data flow) such as sanitization-before-use, parameterized SQL, and bounded operations. Constraints form set $\Phi = \{\phi_1, \phi_2, \ldots, \phi_m\}$, each a boolean predicate. Candidates must satisfy all hard constraints.

\textbf{Hybrid Localization Generation.}
ISAL combines deterministic templates with LLM flexibility: (1) parameterized micro-templates (validation wrappers, prepared statement builders) produce safe skeletons; (2) LLM fills context-specific details under constraint; (3) post-generation rewriting ensures consistency:
\[
s'_k = \text{Rewrite}\big(g_{\theta}(T_k(s,C(s)),\,\Phi)\big).
\]

\textbf{Dual Scoring.}
Each candidate is scored from two perspectives: (a) security score $S_{\mathrm{sec}}(s')$ via re-running frozen Stage I/II detectors, and (b) semantic consistency $S_{\mathrm{sem}}(s')$ via embedding similarity plus static checks (compilation, unit tests):
\[
U(s') = \alpha \cdot S_{\mathrm{sec}}(s') + (1-\alpha)\cdot S_{\mathrm{sem}}(s').
\]

\textbf{Iterative Refinement.}
If no candidate satisfies all constraints, ISAL analyzes failure signals and feeds structured feedback to the LLM for focused regeneration, repeating until iteration budget exhausted.

\begin{algorithm}[t]
\caption{ISAL: Incremental Sequence Analysis and localization}
\label{alg:isair}
\begin{algorithmic}[1]
\Require Vulnerable slice $s$, context $C(s)$, templates $\mathcal{T}$, LLM $g_{\theta}$, constraints $\Phi$, max iterations $I_{\max}$
\Ensure Localized slice $\hat{s}$ or \texttt{FAIL}
\State $\text{IR} \gets \texttt{BuildIR}(s, C(s))$
\State $\Phi \gets \texttt{ExtractConstraints}(\text{IR})$ 
\State $i \gets 0$
\While{$i < I_{\max}$}
    \State $\mathcal{G} \gets [\,]$ 
    \For{each template $T_k \in \mathcal{T}$}
        \State $s'_k \gets \texttt{Rewrite}\!\big(g_{\theta}(T_k(s,C(s)),\,\Phi)\big)$
        \State $\mathcal{G}.\texttt{append}(s'_k)$
    \EndFor
    \State $\hat{s} \gets \texttt{SelectBest}(\mathcal{G}, \Phi, U)$ 
    \If{\texttt{Compile}($\hat{s}$) $\wedge$ \texttt{UnitTest}($\hat{s}$)}
        \State \Return $\hat{s}$
    \Else
        \State $\text{fb} \gets \texttt{AnalyzeFailures}(\mathcal{G}, \Phi)$
        \State $(s,C(s)) \gets \texttt{RefineContext}(s,C(s),\text{fb})$
        \State $i \gets i + 1$
    \EndIf
\EndWhile
\State \Return \texttt{FAIL}
\end{algorithmic}
\end{algorithm}

\section{Evaluation}
\label{sec:evaluation}

To systematically assess the effectiveness of AutoVulnPHP, we design our evaluation to answer the following research questions:
\begin{itemize}
    \item \textbf{RQ1 (Detection Effectiveness):} How does the proposed two-stage cascaded framework compare against state-of-the-art static analysis tools and learning-based models in terms of detection accuracy and false negative reduction?
    \item \textbf{RQ2 (Localization Capability):} To what extent does the ISAL localization module improve localization accuracy and precision compared to unconstrained LLM approaches?
    \item \textbf{RQ3 (Component Contribution):} What are the individual contributions of the AST-guided filtering, risk-biased attention, and constraint-driven generation components to the overall system performance?
    \item \textbf{RQ4 (Practical Impact):} Can AutoVulnPHP effectively discover and aid in reporting previously unknown vulnerabilities in large-scale, real-world PHP applications?
\end{itemize}

\subsection{Experimental Setup}

\textbf{Models.}
We evaluate a two-stage detector followed by a localization module. SIFT-VulMiner performs structure-first detection with a CodeT5 encoder and a GRU classifier. SAFE-VulMiner confirms semantics first with a CodeBERT encoder and a Transformer that uses risk-biased attention. ISAL then carries out iterative, semantic-aware localization with constrained localization result generation and verification. To study interactions between encoders and classifiers, we test all pairings of five encoders and four back ends. The encoders are BERT, CodeBERT, XLNet, GPT2, and CodeT5. The back ends are Transformer, LSTM, GRU, and CNN.

\textbf{Datasets.}
Our primary benchmark is PVts, the PHP Vulnerability test suite \cite{stivalet2015phpvulnsuite}.
We also build PHPVD by adding newly mined positive cases to a larger corpus of real projects, and we use it to assess detection and localization in realistic settings.
For ISAL, we further split vulnerable samples into Injection, URF, and XSS to report localization success by type.

\textbf{Metrics.}
For detection, we evaluate performance using several key metrics: accuracy, precision, F1 score, false positive rate, and false negative rate. Accuracy is the share of samples whose labels are predicted correctly. Precision is the share of samples predicted as vulnerable that are truly vulnerable. The F1 score is the harmonic mean of precision and recall and summarizes the balance between catching vulnerabilities and avoiding false alarms. The false positive rate is the share of benign samples that are incorrectly flagged as vulnerable. The false negative rate is the share of vulnerable samples that the detector fails to flag.

For localization evaluation, we measure the localization rate as the percentage of vulnerable cases that are correctly localized and pass all verification checks.

\begin{table*}[h]
\centering
\caption{Performance of different LLM encoders paired with neural network classifiers.}
\label{tab:llm_nn_combo}
\begin{tabular}{l l c c c c c}
\toprule
\multirow{2}{*}{Encoder} & Classifier & \multicolumn{5}{c}{Metrics} \\
\cmidrule(lr){3-7}
& & ACC & PRE & F1 & FPR & FNR \\
\midrule
\multirow{4}{*}{BERT} 
& Transformer & 79.3\% & 62.8\% & 74.6\% & 27.0\% & 8.1\% \\
& LSTM        & 81.0\% & 62.8\% & 76.0\% & 26.0\% & 3.7\% \\
& GRU         & 81.1\% & 62.9\% & 77.0\% & 27.5\% & 0.7\% \\
& CNN         & 76.0\% & 57.8\% & 71.0\% & 31.5\% & 8.1\% \\
\midrule
\multirow{4}{*}{XLNet} 
& Transformer & 79.0\% & 60.9\% & 73.1\% & 26.6\% & 8.7\% \\
& LSTM        & 80.4\% & 62.0\% & 74.5\% & 25.4\% & 6.7\% \\
& GRU         & 78.4\% & 61.9\% & 71.2\% & 24.1\% & 16.4\% \\
& CNN         & 75.2\% & 54.2\% & 68.8\% & 32.7\% & 5.6\% \\
\midrule
\multirow{4}{*}{GPT-2} 
& Transformer & 75.2\% & 55.0\% & 70.9\% & 35.6\% & 0.1\% \\
& LSTM        & 77.9\% & 58.6\% & 71.1\% & 27.5\% & 9.5\% \\
& GRU         & 77.5\% & 58.7\% & 72.5\% & 30.5\% & 5.1\% \\
& CNN         & 74.3\% & 53.4\% & 69.3\% & 35.9\% & 1.1\% \\
\midrule
\multirow{4}{*}{CodeT5} 
& Transformer & 81.2\% & 65.5\% & 72.5\% & 18.9\% & 18.7\% \\
& LSTM        & 81.2\% & 62.8\% & 75.0\% & 24.1\% & 6.7\% \\
& GRU         & \textbf{83.5\%} & \textbf{65.9\%} & \textbf{79.0\%} & \textbf{23.5\%} & \textbf{1.3\%} \\
& CNN         & 76.1\% & 57.6\% & 69.4\% & 28.9\% & 12.7\% \\
\bottomrule
\end{tabular}
\end{table*}

\textbf{Baselines.}
The following are the baselines used in the experiments.

1) HiddenCPG \cite{wi2022hiddencpg}: A scalable system that uses Code Property Graphs (CPGs) to detect vulnerabilities via subgraph isomorphism, matching known vulnerable patterns in large CPGs with optimization techniques to handle scale.

2) RecurScan \cite{shi2024recurscan}: A method detecting recurring vulnerabilities resilient to code differences, using security patches and symbolic tracking to compare symbolic expressions and selective constraints between target applications and known vulnerabilities.

3) Walden J \cite{walden2014predicting}: A work providing a public dataset of 223 vulnerabilities from three web applications, comparing text mining based and software metrics based vulnerability prediction models, finding text mining models have higher recall.

4) Abunadi I \cite{abunadi2016empirical}: An empirical study exploring cross project vulnerability prediction using machine learning classifiers, comparing five classifiers on a public dataset of PHP web applications to assess their performance in detecting vulnerable components.

5) zero-knowledge LLM: prompt-only, no templates or constraints.

\textbf{Implementation Details.}
Transformer family encoders use lr = $1\times10^{-5}$, batch size 4, dropout 0.3, 10 epochs. For the ISAL framework, Deepseek-R1 is selected as the localization model. Our experiments are conducted on a system with the following setup. For hardware, we use an Intel(R) Xeon(R) Gold 6230 CPU (2.10GHz), 256G memory, and 6 NVIDIA GeForce RTX 3080 GPUs. The software environment includes Ubuntu 20.04, Python 3.10.8, Transformers 4.36.2, PyTorch 2.1.2+cu121, scikit-learn 1.3.2, and numpy 1.26.1. GPU components rely on CUDA 12.2 and cuDNN 8.3.2.

\subsection{Performance Comparison (RQ1 \& RQ2)}

\begin{figure*}[!t]
\centering
\includegraphics[width=0.8\textwidth]{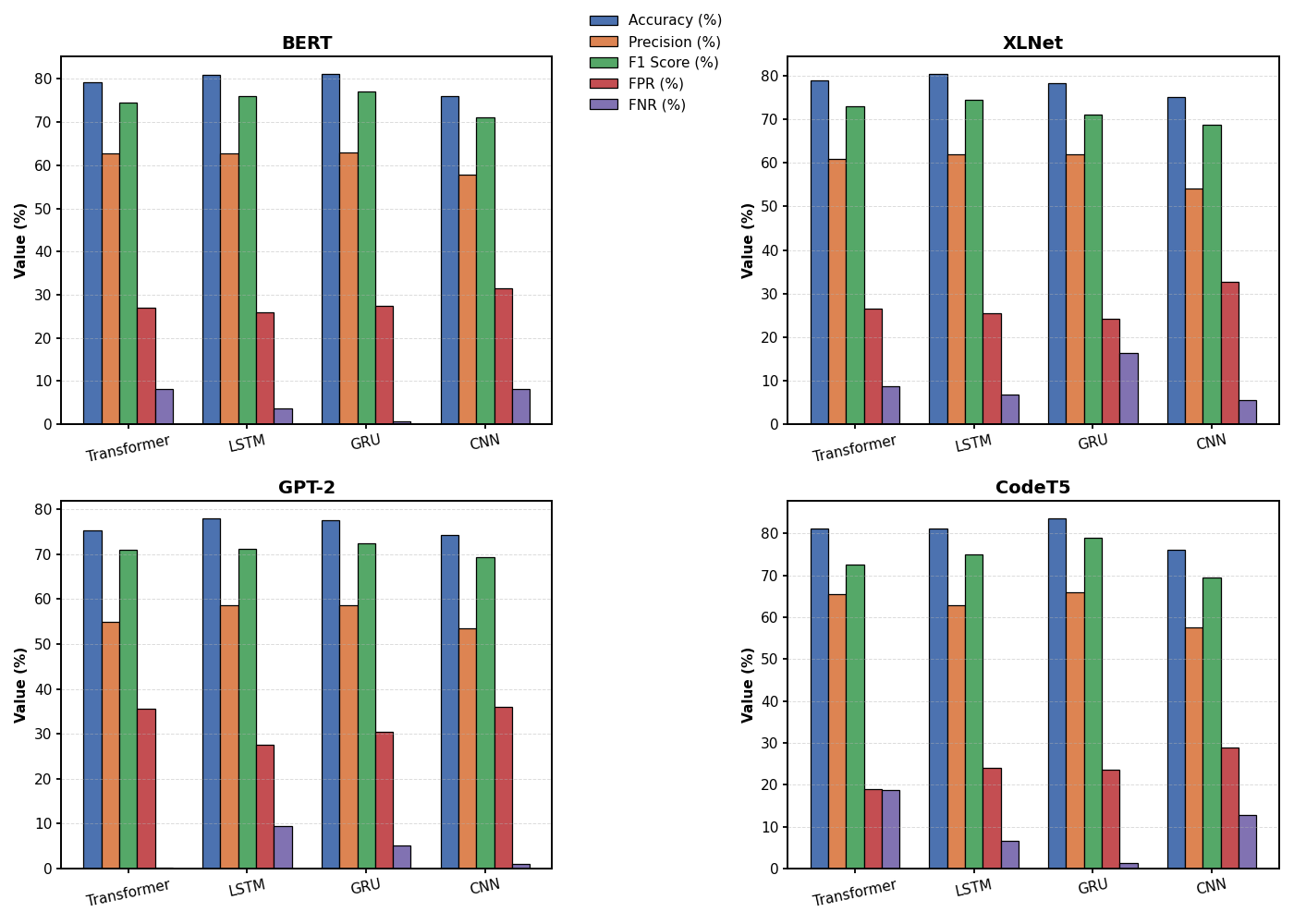}
\caption{Detection performance on PHPVD for different LLM encoder and neural classifier combinations.}
\label{llm_nn_combo_bar}
\end{figure*}

\textbf{Detection Performance.}
Table~\ref{tab:llm_nn_combo} summarizes results on PVts, and Figure~\ref{llm_nn_combo_bar} visualizes them. SIFT-VulMiner (CodeT5+GRU) achieves 83.5\% accuracy with a remarkably low FNR of 1.3\%, demonstrating the efficacy of AST-guided structure in recalling potential threats. Compared with HiddenCPG and RecurScan, the FNR drops sharply from 34.3\% and 14.0\% to 1.3\%. On the larger PHPVD dataset, overall accuracy reaches 89.4\% while the FPR is halved from 23.5\% to 12.9\% via the cascaded design.

\begin{figure}[!t]
\centering
\includegraphics[width=0.48\textwidth]{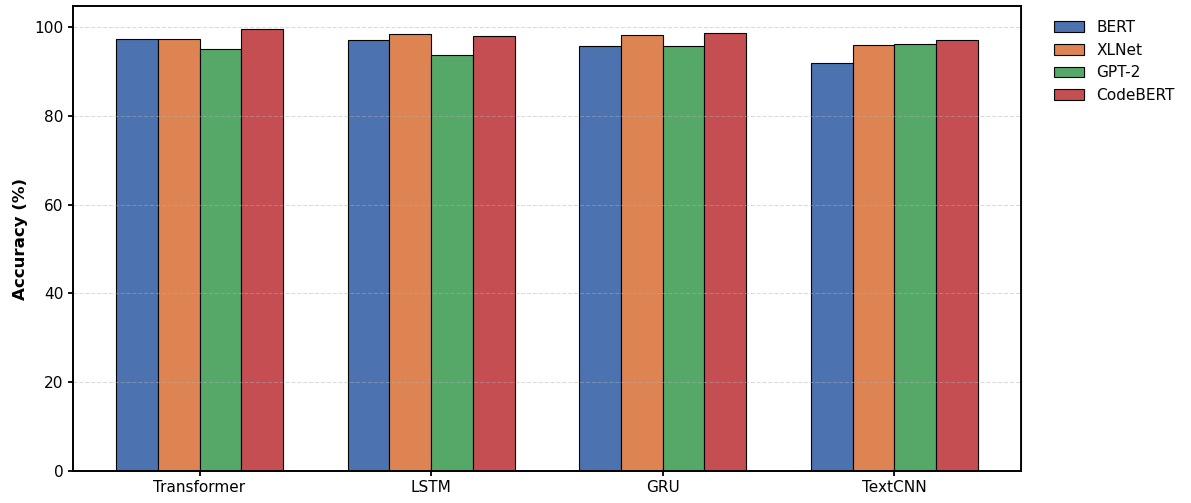}
\caption{Accuracy improvements from data preprocessing and risk-aware attention across LLM and classifier combinations.}
\label{acc_preprocess_attention_bar}
\end{figure}

\begin{table*}[!t]
\centering
\caption{Accuracy before and after data preprocessing and attention adjustment}
\label{tab:acc_preprocess_attention}
\begin{tabular}{lccccc}
\toprule
\textbf{LLM} & \textbf{Transformer} & \textbf{LSTM} & \textbf{GRU} & \textbf{TextCNN} \\
\midrule
BERT & 97.4\% (+18.3) & 97.2\% (+24.3) & 95.8\% (+30.9) & 91.9\% (+15.4) \\
XLNET & 97.4\% (+12.0) & 98.4\% (+16.3) & 98.2\% (+17.8) & 95.9\% (+11.0) \\
GPT2 & 95.1\% (+8.1) & 93.8\% (+9.2) & 95.8\% (+8.8) & 96.3\% (+7.0) \\
CodeBERT & 99.7\% (+13.3) & 97.9\% (+21.5) & 98.7\% (+19.1) & 97.2\% (+14.8) \\
\bottomrule
\end{tabular}
\end{table*}

\begin{table}[t]
\centering
\caption{Effect of removing AST in SIFT-VulMiner.}
\label{tab:lat_ast_ablation}
\begin{tabular}{l c c c}
\toprule
Variant & ACC & FPR & FNR \\
\midrule
Full (AST + Enc + GRU) & 83.5\% & 23.5\% & \textbf{1.3\%} \\
w/o AST (raw code only) & 79.2\% & 28.9\% & 7.8\% \\
\bottomrule
\end{tabular}
\end{table}

\begin{table}[t]
\centering
\caption{Localization rate (\%) comparison between zero-knowledge LLM and ISAL.}
\label{tab:isair_repair}
\begin{tabular}{l c c c c}
\toprule
Method & Injection & URF & XSS & Total \\
\midrule
Zero-knowledge LLM & 66.89 & 24.71 & 87.36 & 65.32 \\
ISAL               & \textbf{84.97} & \textbf{48.26} & \textbf{94.94} & \textbf{80.95} \\
\bottomrule
\end{tabular}
\end{table}

\begin{table}[t]
\centering
\caption{ISAL ablation on overall localization rate (\%).}
\label{tab:isair_ablation}
\begin{tabular}{l c}
\toprule
Variant & Localization rate \\
\midrule
Full ISAL                & \textbf{80.95} \\
w/o Iterative Refinement  & 77.12 \\
w/o Constraint Set $\Phi$ & 69.34 \\
w/o Template Library $\mathcal{T}$ & 71.05 \\
Prompt only LLM           & 65.32 \\
\bottomrule
\end{tabular}
\end{table}

\begin{figure}[!t]  
\centering
\includegraphics[width=1.0\linewidth]{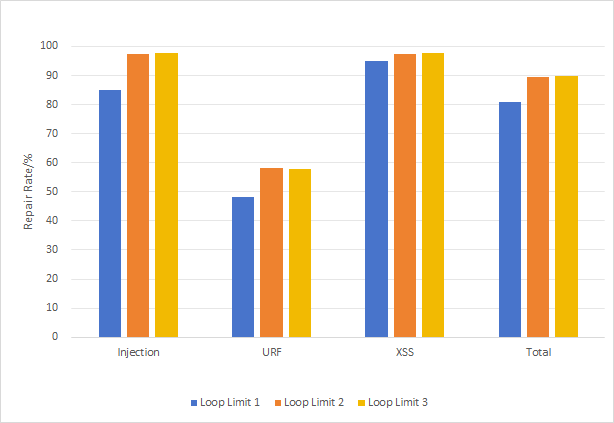}
\caption{Localization rate and number of loops.} 
\label{Repair Rate loop}
\end{figure}

\textbf{Localization Performance.}
Table~\ref{tab:isair_repair} reports the localization rates. ISAL achieves an overall localization rate of 80.95\%, significantly outperforming the zero-knowledge LLM baseline (65.32\%). The improvement is most consistent in Injection (84.97\%) and XSS (94.94\%) cases, where template-guided constraints effectively prevent common escaping errors.

\textbf{Failure Analysis.}
Despite ISAL's high success rate, approximately 19\% of cases remain unsolved. We identified three primary challenges limiting the success rate:
\begin{itemize}
    \item \textbf{Complex Inter-procedural Context:} Some vulnerabilities originate from data flows spanning multiple files or classes. Since ISAL currently operates on retrieved code slices, it may lack the full context required to generate a precise localization result without breaking dependencies.
    \item \textbf{Constraint Conflicts:} In 7\% of failed cases, the extracted security constraints conflicted with functional requirements (e.g., sanitizing an input that required raw HTML formatting), causing the verification step to reject the inaccurate localization result to preserve functionality.
    \item \textbf{LLM Hallucination on Dynamic Features:} For highly dynamic PHP constructs (e.g., variable variables or `call\_user\_func`), the LLM occasionally failed to adhere to the strict syntax templates, generating code that was secure but syntactically invalid in the target PHP version.
\end{itemize}

\subsection{Ablation Study (RQ3)}
\label{subsec:ablation}

\textbf{Two-Stage Detection Components.}
Removing AST linearization in Stage I increases the FNR widely (Table~\ref{tab:lat_ast_ablation}), confirming the need for structural modeling. In Stage II, disabling risk-aware attention leads to accuracy drops of up to 23.3\% (CodeBERT+Transformer), verifying that the bias term $\mathbf{B}$ is essential for focusing on sparse vulnerability signals in large codebases.

\textbf{ISAL Components.}
Table~\ref{tab:isair_ablation} reports an ablation study of ISAL, where we evaluate the contribution of the template library $\mathcal{T}$, the constraint set $\Phi$, and the iterative refinement loop to the overall localization rate. Full ISAL achieves a localization rate of 80.95\%, which drops to 77.12\% without iterative refinement, 69.34\% without the constraint set, 71.05\% without the template library, and 65.32\% for the prompt-only LLM.
The constraint set delivers the largest security gains. It enforces required checks and steers patches toward the root cause of each flaw. The template library improves stability by offering safe patch patterns that the model completes with context. The iterative loop mainly rescues difficult cases that fail on the first attempt. As shown in Figure~\ref{Repair Rate loop} and supported by the ablation results, performance saturates after the second iteration; additional rounds add cost with little benefit. We therefore use two iterations in all experiments.

\subsection{Real-World Vulnerability Discovery (RQ4)}

To validate the practical utility of AutoVulnPHP beyond benchmarks, we deployed the framework in a wild mining campaign.
\textbf{Setup.} We crawled massive open-source PHP repositories from GitHub and source code hosting platforms. The code was preprocessed and fed into the AutoVulnPHP pipeline.
\textbf{Results.} The system identified a total of 429 previously unknown vulnerabilities. Of these, 351 were verified as genuine high-risk issues and have been assigned CVE identifiers following our responsible disclosure.
\textbf{Impact.} This large-scale discovery demonstrates that AutoVulnPHP generalizes well to diverse coding styles and project structures. The integration of automated localization results also streamlined the disclosure process, allowing us to provide maintainers with precise localization reports alongside vulnerability reports.

\section{Conclusion}
\label{sec:conclusion}
AutoVulnPHP provides a complete PHP vulnerability framework that couples precise detection with practical localization. The two-stage VulMiner pipeline first uses SIFT-VulMiner to extract structural signals and then applies SAFE-VulMiner for semantic confirmation, while ISAL performs template-guided, constraint-aware, and iteratively refined localization result generation and verification. On public benchmarks and the PHPVD dataset (26{,}614 real-world PHP files with 5.2M lines of code), AutoVulnPHP achieves 99.7\% detection accuracy and an 81\% localization success rate. Using this framework, we discovered 429 previously unknown vulnerabilities, of which 351 have received CVE identifiers. Together with the obfuscation-aware PHPVD dataset, these results show that combining AST-guided structure, semantic analysis, and constraint-driven localization can sharply reduce missed vulnerabilities while maintaining practical, high-quality localization outcomes.

Future work includes extending the framework beyond PHP to other languages, adding stronger verification, and incorporating developer feedback to improve guidance and interpretability. We also plan to apply the same detection and localization pipeline to broader LLM-based program analysis tasks such as secure refactoring and exploit prevention.

\bibliographystyle{ACM-Reference-Format}
\bibliography{sample-base}

@String{Computing = "Computing" }

@String{Computer = "{IEEE} Computer" }

@String{Springer = "Springer-Verlag" }

@techreport{synopsys2024,
  author       = {{Synopsys Technology}},
  title        = {2024 Open Source Security and Risk Analysis Report},
  type         = {Online report},
  institution  = {Synopsys Black Duck Audit Services},
  year         = {2024},
  url          = {https://www.blackduck.com/resources/analyst-reports/open-source-security-risk-analysis/},
  note         = {Electronic resource accessed in 2025}
}

@misc{wordpress2025,
  author        = {{WordPress}},
  title         = {WordPress [EB]},
  howpublished  = {\url{https://wordpress.org/download}},
  year          = {2025},
  note          = {Electronic resource}
}

@misc{facebook2024,
  author        = {{Facebook}},
  title         = {Facebook [EB]},
  howpublished  = {\url{https://www.facebook.com}},
  year          = {2024},
  note          = {Electronic resource}
}

@misc{w3techs2025,
  author        = {{W3Techs}},
  title         = {Usage Statistics and Market Share of PHP for Websites [EB]},
  howpublished  = {\url{https://w3techs.com/technologies/details/pl-php}},
  year          = {2025},
  note          = {Electronic resource}
}

@inproceedings{shi2024recurscan,
  title={Recurscan: Detecting recurring vulnerabilities in php web applications},
  author={Shi, Youkun and Zhang, Yuan and Bai, Tianhao and Zhang, Lei and Tan, Xin and Yang, Min},
  booktitle={Proceedings of the ACM Web Conference 2024},
  pages={1746--1755},
  year={2024}
}

@inproceedings{saccente2019project,
  title={Project achilles: A prototype tool for static method-level vulnerability detection of java source code using a recurrent neural network},
  author={Saccente, Nicholas and Dehlinger, Josh and Deng, Lin and Chakraborty, Suranjan and Xiong, Yin},
  booktitle={2019 34th IEEE/ACM International Conference on Automated Software Engineering Workshop (ASEW)},
  pages={114--121},
  year={2019},
  organization={IEEE}
}

@inproceedings{feng2020efficient,
  title={Efficient vulnerability detection based on abstract syntax tree and deep learning},
  author={Feng, Hantao and Fu, Xiaotong and Sun, Hongyu and Wang, He and Zhang, Yuqing},
  booktitle={IEEE INFOCOM 2020-IEEE Conference on Computer Communications Workshops (INFOCOM WKSHPS)},
  pages={722--727},
  year={2020},
  organization={IEEE}
}

@article{bengio2021deep,
  title={Deep learning for AI},
  author={Bengio, Yoshua and Lecun, Yann and Hinton, Geoffrey},
  journal={Communications of the ACM},
  volume={64},
  number={7},
  pages={58--65},
  year={2021},
  publisher={ACM New York, NY, USA}
}

@misc{stivalet2015phpvulnsuite,
  author       = {{PHP-Vulnerability-test-suite} team},
  title        = {PHP-Vulnerability-test-suite},
  year         = {2015},
  publisher    = {GitHub},
  url          = {https://github.com/stivalet/PHP-Vulnerability-test-suite}
}

@inproceedings{walden2014predicting,
  title={Predicting vulnerable components: Software metrics vs text mining},
  author={Walden, James and Stuckman, Jeff and Scandariato, Riccardo},
  booktitle={2014 IEEE 25th international symposium on software reliability engineering},
  pages={23--33},
  year={2014},
  organization={IEEE}
}

@article{abunadi2016empirical,
  title={An empirical investigation of security vulnerabilities within web applications.},
  author={Abunadi, Ibrahim and Alenezi, Mamdouh},
  journal={J. Univers. Comput. Sci.},
  volume={22},
  number={4},
  pages={537--551},
  year={2016}
}

@inproceedings{cha2012unleashing,
  title={Unleashing mayhem on binary code},
  author={Cha, Sang Kil and Avgerinos, Thanassis and Rebert, Alexandre and Brumley, David},
  booktitle={2012 IEEE Symposium on Security and Privacy},
  pages={380--394},
  year={2012},
  organization={IEEE}
}

@inproceedings{wi2022hiddencpg,
  title={HiddenCPG: large-scale vulnerable clone detection using subgraph isomorphism of code property graphs},
  author={Wi, Seongil and Woo, Sijae and Whang, Joyce Jiyoung and Son, Sooel},
  booktitle={Proceedings of the ACM web conference 2022},
  pages={755--766},
  year={2022}
}

@misc{CVEProgram,
    author       = {{CVE Program}},
    title        = {CVE Program Website[EB/OL]},
    url          = {https://cve.mitre.org/},
    note         = {Legacy website, now redirected to WWW.CVE.ORG; Mission: To identify, define, and catalog publicly disclosed cybersecurity vulnerabilities. Total CVE Records: 285,863 (as of 2025-06-04).},
    year         = {2025}
}

@inproceedings{cheatham1964syntax,
  title={Syntax-directed compiling},
  author={Cheatham Jr, Thomas E and Sattley, Kirk},
  booktitle={Proceedings of the April 21-23, 1964, spring joint computer conference},
  pages={31--57},
  year={1964}
}

@inproceedings{yamaguchi2014modeling,
  title={Modeling and discovering vulnerabilities with code property graphs},
  author={Yamaguchi, Fabian and Golde, Nico and Arp, Daniel and Rieck, Konrad},
  booktitle={2014 IEEE symposium on security and privacy},
  pages={590--604},
  year={2014},
  organization={IEEE}
}

@article{cadar2013symbolic,
  title={Symbolic execution for software testing: three decades later},
  author={Cadar, Cristian and Sen, Koushik},
  journal={Communications of the ACM},
  volume={56},
  number={2},
  pages={82--90},
  year={2013},
  publisher={ACM New York, NY, USA}
}

@article{hu2025sok,
  title={SoK: Automated Vulnerability Repair: Methods, Tools, and Assessments},
  author={Hu, Yiwei and Li, Zhen and Shu, Kedie and Guan, Shenghua and Zou, Deqing and Xu, Shouhuai and Yuan, Bin and Jin, Hai},
  journal={arXiv preprint arXiv:2506.11697},
  year={2025}
}

@inproceedings{xie2006static,
  title={Static Detection of Security Vulnerabilities in Scripting Languages.},
  author={Xie, Yichen and Aiken, Alex},
  booktitle={USENIX Security Symposium},
  volume={15},
  pages={179--192},
  year={2006}
}

@article{wei2022chain,
  title={Chain-of-thought prompting elicits reasoning in large language models},
  author={Wei, Jason and Wang, Xuezhi and Schuurmans, Dale and Bosma, Maarten and Xia, Fei and Chi, Ed and Le, Quoc V and Zhou, Denny and others},
  journal={Advances in neural information processing systems},
  volume={35},
  pages={24824--24837},
  year={2022}
}

@article{cho2014learning,
  title={Learning phrase representations using RNN encoder-decoder for statistical machine translation},
  author={Cho, Kyunghyun and Van Merri{\"e}nboer, Bart and Gulcehre, Caglar and Bahdanau, Dzmitry and Bougares, Fethi and Schwenk, Holger and Bengio, Yoshua},
  journal={arXiv preprint arXiv:1406.1078},
  year={2014}
}

@article{vaswani2017attention,
  title={Attention is all you need},
  author={Vaswani, Ashish and Shazeer, Noam and Parmar, Niki and Uszkoreit, Jakob and Jones, Llion and Gomez, Aidan N and Kaiser, {\L}ukasz and Polosukhin, Illia},
  journal={Advances in neural information processing systems},
  volume={30},
  year={2017}
}

@inproceedings{li2025tuni,
  title={Tuni: A textual unimodal detector for identity inference in clip models},
  author={Li, Songze and Cheng, Ruoxi and Jia, Xiaojun},
  booktitle={Proceedings of the Sixth Workshop on Privacy in Natural Language Processing},
  pages={1--13},
  year={2025}
}

@inproceedings{cheng2025pbi,
  title={Pbi-attack: Prior-guided bimodal interactive black-box jailbreak attack for toxicity maximization},
  author={Cheng, Ruoxi and Ding, Yizhong and Cao, Shuirong and Duan, Ranjie and Jia, Xiaoshuang and Yuan, Shaowei and Qin, Simeng and Wang, Zhiqiang and Jia, Xiaojun},
  booktitle={Proceedings of the 2025 Conference on Empirical Methods in Natural Language Processing},
  pages={609--628},
  year={2025}
}

@article{chengpbi,
  title={PBI-Attack: Prior-Guided Bimodal Interactive Black-Box Jailbreak Attack for Toxicity Maximization},
  author={Cheng, Ruoxi and Ding, Yizhong and Cao, Shuirong and Yuan, Shaowei and Duan, Ranjie and Jia, Xiaoshuang and Wang, Zhiqiang and Jia, Xiaojun}
}

@inproceedings{pei2025selfprompt,
  title={SelfPrompt: Autonomously Evaluating LLM Robustness via Domain-Constrained Knowledge Guidelines and Refined Adversarial Prompts},
  author={Pei, Aihua and Yang, Zehua and Zhu, Shunan and Cheng, Ruoxi and Jia, Ju},
  booktitle={COLING},
  year={2025}
}

@inproceedings{cao2025agr,
  title={Agr: Age group fairness reward for bias mitigation in llms},
  author={Cao, Shuirong and Cheng, Ruoxi and Wang, Zhiqiang},
  booktitle={ICASSP 2025-2025 IEEE International Conference on Acoustics, Speech and Signal Processing (ICASSP)},
  pages={1--5},
  year={2025},
  organization={IEEE}
}

@article{duan2025oyster,
  title={Oyster-I: Beyond Refusal--Constructive Safety Alignment for Responsible Language Models},
  author={Duan, Ranjie and Liu, Jiexi and Jia, Xiaojun and Zhao, Shiji and Cheng, Ruoxi and Wang, Fengxiang and Wei, Cheng and Xie, Yong and Liu, Chang and Li, Defeng and others},
  journal={arXiv preprint arXiv:2509.01909},
  year={2025}
}

@article{cheng2025inverse,
  title={Inverse reinforcement learning with dynamic reward scaling for llm alignment},
  author={Cheng, Ruoxi and Ma, Haoxuan and Wang, Weixin and Duan, Ranjie and Liu, Jiexi and Jia, Xiaoshuang and Qin, Simeng and Cao, Xiaochun and Liu, Yang and Jia, Xiaojun},
  journal={arXiv preprint arXiv:2503.18991},
  year={2025}
}

@inproceedings{cheng2025gibberish,
  title={Gibberish is all you need for membership inference detection in contrastive language-audio pretraining},
  author={Cheng, Ruoxi and Ding, Yizhong and Cao, Shuirong and Wang, Zhiqiang},
  booktitle={Proceedings of the 2025 International Conference on Multimedia Retrieval},
  pages={108--116},
  year={2025}
}

@article{cheng2024gibberish,
  title={Gibberish is All You Need for Membership Inference Detection in Contrastive Language-Audio Pretraining},
  author={Cheng, Ruoxi and Ding, Yizhong and Cao, Shuirong and Shao, Shitong and Wang, Zhiqiang},
  journal={arXiv preprint arXiv:2410.18371},
  year={2024}
}

@article{zhao2025strata,
  title={Strata-sword: A hierarchical safety evaluation towards llms based on reasoning complexity of jailbreak instructions},
  author={Zhao, Shiji and Duan, Ranjie and Liu, Jiexi and Jia, Xiaojun and Wang, Fengxiang and Wei, Cheng and Cheng, Ruoxi and Xie, Yong and Liu, Chang and Guo, Qing and others},
  journal={arXiv preprint arXiv:2509.01444},
  year={2025}
}

@article{li2025clipscore,
  title={L-clipscore: a lightweight embedding-based captioning metric for evaluating and training},
  author={Li, Li and Peng, Yingzhe and Yang, Xu and Cheng, Ruoxi and Xu, Haiyang and Yan, Ming and Huang, Fei},
  journal={arXiv preprint arXiv:2507.08710},
  year={2025}
}

@article{cheng2025ecoalign,
  title={EcoAlign: An Economically Rational Framework for Efficient LVLM Alignment},
  author={Cheng, Ruoxi and Ma, Haoxuan and Ma, Teng and Zhang, Hongyi},
  journal={arXiv preprint arXiv:2511.11301},
  year={2025}
}

@article{teng2024heuristic,
  title={Heuristic-induced multimodal risk distribution jailbreak attack for multimodal large language models},
  author={Teng, Ma and Xiaojun, Jia and Ranjie, Duan and Xinfeng, Li and Yihao, Huang and Xiaoshuang, Jia and Zhixuan, Chu and Wenqi, Ren},
  journal={arXiv preprint arXiv:2412.05934},
  year={2024}
}

@article{jia2025omnisafebench,
  title={OmniSafeBench-MM: A Unified Benchmark and Toolbox for Multimodal Jailbreak Attack-Defense Evaluation},
  author={Jia, Xiaojun and Liao, Jie and Guo, Qi and Ma, Teng and Qin, Simeng and Duan, Ranjie and Li, Tianlin and Huang, Yihao and Zeng, Zhitao and Wu, Dongxian and others},
  journal={arXiv preprint arXiv:2512.06589},
  year={2025}
}

@inproceedings{cao2024agr,
  title={AGR: Age Group fairness Reward for Bias Mitigation in LLMs},
  author={Cao, Shuirong and Cheng, Ruoxi and others},
  booktitle={Pluralistic Alignment Workshop at NeurIPS 2024}
}

@inproceedings{cheng2025talk,
  title={Talk the talk, debate the bias: Llm alignment via role-play rumble},
  author={Cheng, Ruoxi and Wang, Zhiqiang and Yuan, Shaowei and Ding, Yizhong and Zhang, Rui},
  booktitle={International Conference on Intelligent Computing},
  pages={201--211},
  year={2025},
  organization={Springer}
}

@article{cheng2024deceiving,
  title={Deceiving to enlighten: Coaxing llms to self-reflection for enhanced bias detection and mitigation},
  author={Cheng, Ruoxi and Ma, Haoxuan and Cao, Shuirong},
  journal={arXiv preprint arXiv:2404.10160},
  year={2024}
}

@article{cheng2024rlrf,
  title={Rlrf: Reinforcement learning from reflection through debates as feedback for bias mitigation in llms},
  author={Cheng, Ruoxi and Ma, Haoxuan and Cao, Shuirong and Shi, Tianyu},
  journal={arXiv preprint arXiv:2404.10160},
  year={2024}
}

@article{ge2025mrfd,
  title={Mrfd: Multi-region fusion decoding with self-consistency for mitigating hallucinations in lvlms},
  author={Ge, Haonan and Wang, Yiwei and Yang, Ming-Hsuan and Cai, Yujun},
  journal={arXiv preprint arXiv:2508.10264},
  year={2025}
}

@article{ge2025framemind,
  title={FrameMind: Frame-Interleaved Video Reasoning via Reinforcement Learning},
  author={Ge, Haonan and Wang, Yiwei and Chang, Kai-Wei and Wu, Hang and Cai, Yujun},
  journal={arXiv preprint arXiv:2509.24008},
  year={2025}
}

@article{yang2024lever,
  title={Lever LM: configuring in-context sequence to lever large vision language models},
  author={Yang, Xu and Peng, Yingzhe and Ma, Haoxuan and Xu, Shuo and Zhang, Chi and Han, Yucheng and Zhang, Hanwang},
  journal={Advances in Neural Information Processing Systems},
  volume={37},
  pages={100341--100368},
  year={2024}
}

@misc{wu2025sokunderstandingnewsecurity,
      title={SoK: Understanding (New) Security Issues Across AI4Code Use Cases}, 
      author={Qilong Wu and Taoran Li and Tianyang Zhou and Varun Chandrasekaran},
      year={2025},
      eprint={2512.18456},
      archivePrefix={arXiv},
      primaryClass={cs.CR},
      url={https://arxiv.org/abs/2512.18456}, 
}

@article{jiang2025never,
  title={Never compromise to vulnerabilities: A comprehensive survey on ai governance},
  author={Jiang, Yuchu and Zhao, Jian and Yuan, Yuchen and Zhang, Tianle and Huang, Yao and Zhang, Yanghao and Wang, Yan and Li, Yanshu and Guo, Xizhong and Zhao, Yusheng and others},
  journal={arXiv preprint arXiv:2508.08789},
  year={2025}
}

@inproceedings{cheng2025speaker,
  title={Speaker Inference Detection Using Only Text},
  author={Cheng, Ruoxi and Ding, Yizhong and Yuan, Shaowei and Wang, Zhiqiang},
  booktitle={International Conference on Information and Communications Security},
  pages={277--294},
  year={2025},
  organization={Springer}
}

@inproceedings{cheng2025usmid,
  title={USMID: A Unimodal Speaker-Level Membership Inference Detector for Contrastive Pretraining},
  author={Cheng, Ruoxi and Ding, Yizhong and Cao, Shuirong and Shao, Shitong and Wang, Zhiqiang},
  booktitle={2025 IEEE International Conference on Acoustics, Speech, and Signal Processing Workshops (ICASSPW)},
  pages={1--5},
  year={2025},
  organization={IEEE}
}

@article{cheng2024unimodal,
  title={A Unimodal Speaker-Level Membership Inference Detector for Contrastive Pretraining},
  author={Cheng, Ruoxi and Ding, Yizhong and Cao, Shuirong and Shao, Shitong and Wang, Zhiqiang},
  journal={arXiv e-prints},
  pages={arXiv--2410},
  year={2024}
}

@inproceedings{wang2025introspective,
  title={Introspective reward modeling via inverse reinforcement learning for llm alignment},
  author={Wang, Zhiqiang and Cheng, Ruoxi and Yuan, Shaowei and Ding, Yizhong and Zhang, Rui},
  booktitle={International Conference on Intelligent Computing},
  pages={233--244},
  year={2025},
  organization={Springer}
}

@article{cheng2025hair,
  title={Hair: Hardness-aware inverse reinforcement learning with introspective reasoning for llm alignment},
  author={Cheng, Ruoxi and Ma, Haoxuan and Wang, Weixin},
  journal={arXiv e-prints},
  pages={arXiv--2503},
  year={2025}
}

@article{li2024identity,
  title={Identity inference from clip models using only textual data},
  author={Li, Songze and Cheng, Ruoxi and Jia, Xiaojun},
  journal={arXiv e-prints},
  pages={arXiv--2405},
  year={2024}
}

@article{wei2025dual,
  title={Dual-Priv Pruning: Efficient Differential Private Fine-Tuning in Multimodal Large Language Models},
  author={Wei, Qianshan and Li, Jiaqi and You, Zihan and Zhan, Yi and Li, Kecen and Wu, Jialin and Liu, Xinfeng Li Hengjun and Yu, Yi and Cao, Bin and Xu, Yiwen and others},
  journal={arXiv preprint arXiv:2506.07077},
  year={2025}
}

@article{li2024single,
  title={Single image unlearning: Efficient machine unlearning in multimodal large language models},
  author={Li, Jiaqi and Wei, Qianshan and Zhang, Chuanyi and Qi, Guilin and Du, Miaozeng and Chen, Yongrui and Bi, Sheng and Liu, Fan},
  journal={Advances in Neural Information Processing Systems},
  volume={37},
  pages={35414--35453},
  year={2024}
}

@article{feng2020codebert,
  title={Codebert: A pre-trained model for program-ming and natural languages},
  author={Feng, Z},
  journal={arXiv preprint arXiv:2002.08155},
  year={2020}
}

@article{wang2021codet5,
  title={Codet5: Identifier-aware unified pre-trained encoder-decoder models for code understanding and generation},
  author={Wang, Yue and Wang, Weishi and Joty, Shafiq and Hoi, Steven CH},
  journal={arXiv preprint arXiv:2109.00859},
  year={2021}
}

@inproceedings{bohme2016coverage,
  title={Coverage-based greybox fuzzing as markov chain},
  author={B{\"o}hme, Marcel and Pham, Van-Thuan and Roychoudhury, Abhik},
  booktitle={Proceedings of the 2016 ACM SIGSAC Conference on Computer and Communications Security},
  pages={1032--1043},
  year={2016}
}

@article{chipounov2012s2e,
  title={The S2E platform: Design, implementation, and applications},
  author={Chipounov, Vitaly and Kuznetsov, Volodymyr and Candea, George},
  journal={ACM Transactions on Computer Systems (TOCS)},
  volume={30},
  number={1},
  pages={1--49},
  year={2012},
  publisher={ACM New York, NY, USA}
}

@inproceedings{ispoglou2020fuzzgen,
  title={$\{$FuzzGen$\}$: Automatic fuzzer generation},
  author={Ispoglou, Kyriakos and Austin, Daniel and Mohan, Vishwath and Payer, Mathias},
  booktitle={29th USENIX Security Symposium (USENIX Security 20)},
  pages={2271--2287},
  year={2020}
}

@inproceedings{rabheru2022hybrid,
  title={A hybrid graph neural network approach for detecting PHP vulnerabilities},
  author={Rabheru, Rishi and Hanif, Hazim and Maffeis, Sergio},
  booktitle={2022 IEEE Conference on Dependable and Secure Computing (DSC)},
  pages={1--9},
  year={2022},
  organization={IEEE}
}

@inproceedings{risse2024uncovering,
  title={Uncovering the limits of machine learning for automatic vulnerability detection},
  author={Risse, Niklas and B{\"o}hme, Marcel},
  booktitle={33rd USENIX Security Symposium (USENIX Security 24)},
  pages={4247--4264},
  year={2024}
}

@inproceedings{siddiq2021sqlifix,
  title={SQLIFIX: Learning based approach to fix SQL injection vulnerabilities in source code},
  author={Siddiq, Mohammed Latif and Jahin, Md Rezwanur Rahman and Islam, Mohammad Rafid Ul and Shahriyar, Rifat and Iqbal, Anindya},
  booktitle={2021 IEEE International Conference on Software Analysis, Evolution and Reengineering (SANER)},
  pages={354--364},
  year={2021},
  organization={IEEE}
}

@inproceedings{kalouptsoglou2024vulnerability,
  title={Vulnerability prediction using pre-trained models: An empirical evaluation},
  author={Kalouptsoglou, Ilias and Siavvas, Miltiadis and Ampatzoglou, Apostolos and Kehagias, Dionysios and Chatzigeorgiou, Alexander},
  booktitle={2024 32nd International Conference on Modeling, Analysis and Simulation of Computer and Telecommunication Systems (MASCOTS)},
  pages={1--6},
  year={2024},
  organization={IEEE}
}

@inproceedings{kulsum2024case,
  title={A case study of llm for automated vulnerability repair: Assessing impact of reasoning and patch validation feedback},
  author={Kulsum, Ummay and Zhu, Haotian and Xu, Bowen and d'Amorim, Marcelo},
  booktitle={Proceedings of the 1st ACM International Conference on AI-Powered Software},
  pages={103--111},
  year={2024}
}

\clearpage
\appendix

\section{Prompt template}

The following is the CoT prompt template employed by ISAL.

\begin{verbatim}
<Input>PHP code and line number
You are a PHP security analysis engine. Input PHP code and line number, and 
output the vulnerability classification result. Strictly follow the following
rules:
Input format:
{php_code: "complete PHP code", line: line number}
Thinking and analysis process:
Quickly parse the code logic and data flow.
Focus on checking the code at the specified line number to determine if there
are security vulnerabilities.
Determine the vulnerability type according to common PHP vulnerability types
(such as SQL injection, XSS, command injection, file inclusion vulnerability,
CSRF, path traversal, etc.).
Analyze the cause of the vulnerability.
Finally, output in the following JSON format:
{
"vulnerability type": "vulnerability name",
"cause analysis": "briefly describe the cause of the vulnerability",
"involved line numbers": "the line number where the vulnerability is located"
}
\end{verbatim}

\newpage
\clearpage

\section{Case Study}
\label{subsec:case-study}

\begin{figure*}[!t]
\centering
\includegraphics[width=1.0\textwidth]{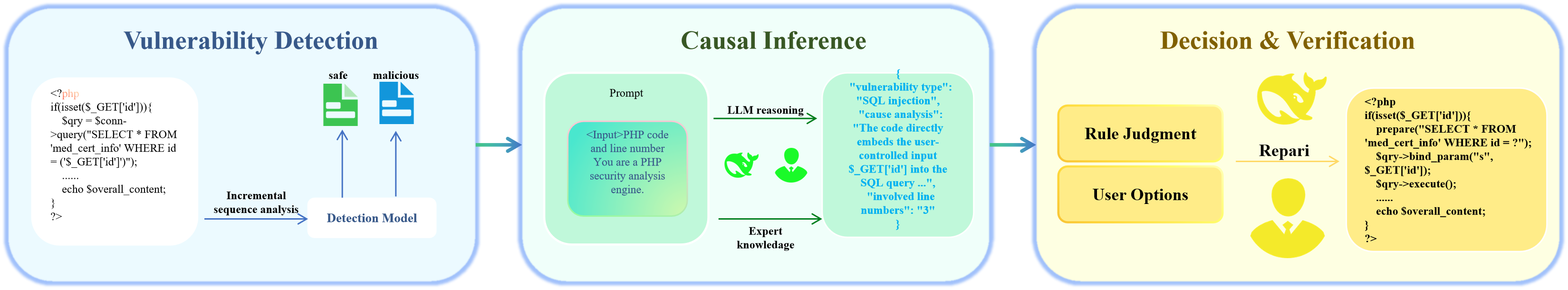}
\caption{Case study: localizing vulnerabilities through ISAL template-guided localization and iterative refinement.}
\label{casestudy}
\end{figure*}

Figure ~\ref{casestudy} illustrates two representative PHP vulnerabilities from our dataset and the corresponding localization results generated by ISAL.

Hidden Command Injection.
A PHP snippet builds a shell command from unsanitized input:
\begin{lstlisting}[language=PHP,basicstyle=\ttfamily\footnotesize,breaklines=true,showstringspaces=false]
$cmd = "tar -czf ".$_GET['file'].".tar.gz ".$_GET['path'];
system($cmd);
\end{lstlisting}

SIFT-VulMiner finds the taint path (\texttt{\$\_GET} $\rightarrow$ \texttt{system()}); ISAL pinpoints the vulnerable code segments and provides guidance with sanitization and safe wrapper suggestions:
\begin{lstlisting}[language=PHP,basicstyle=\ttfamily\footnotesize,breaklines=true,showstringspaces=false]
$path = sanitize_path($_GET['path']);
$file = sanitize_filename($_GET['file']);
$cmd  = escapeshellcmd("tar -czf {$file}.tar.gz {$path}");
exec($cmd);
\end{lstlisting}
Localization result accurately identifies the vulnerable segments and the suggested safeguards pass tests.

SQL Injection in Authentication.
risk-biased attention in SAFE-VulMiner highlights the suspicious \texttt{"SELECT ... WHERE username=\allowbreak'\$\textunderscore POST[user]'"} clause. ISAL accurately localizes the vulnerable clause and provides template-based guidance for secure statement construction with context-specific details.

\section{Training Curves}

\begin{figure*}[htbp]
\centering
\includegraphics[width=1.0\linewidth]{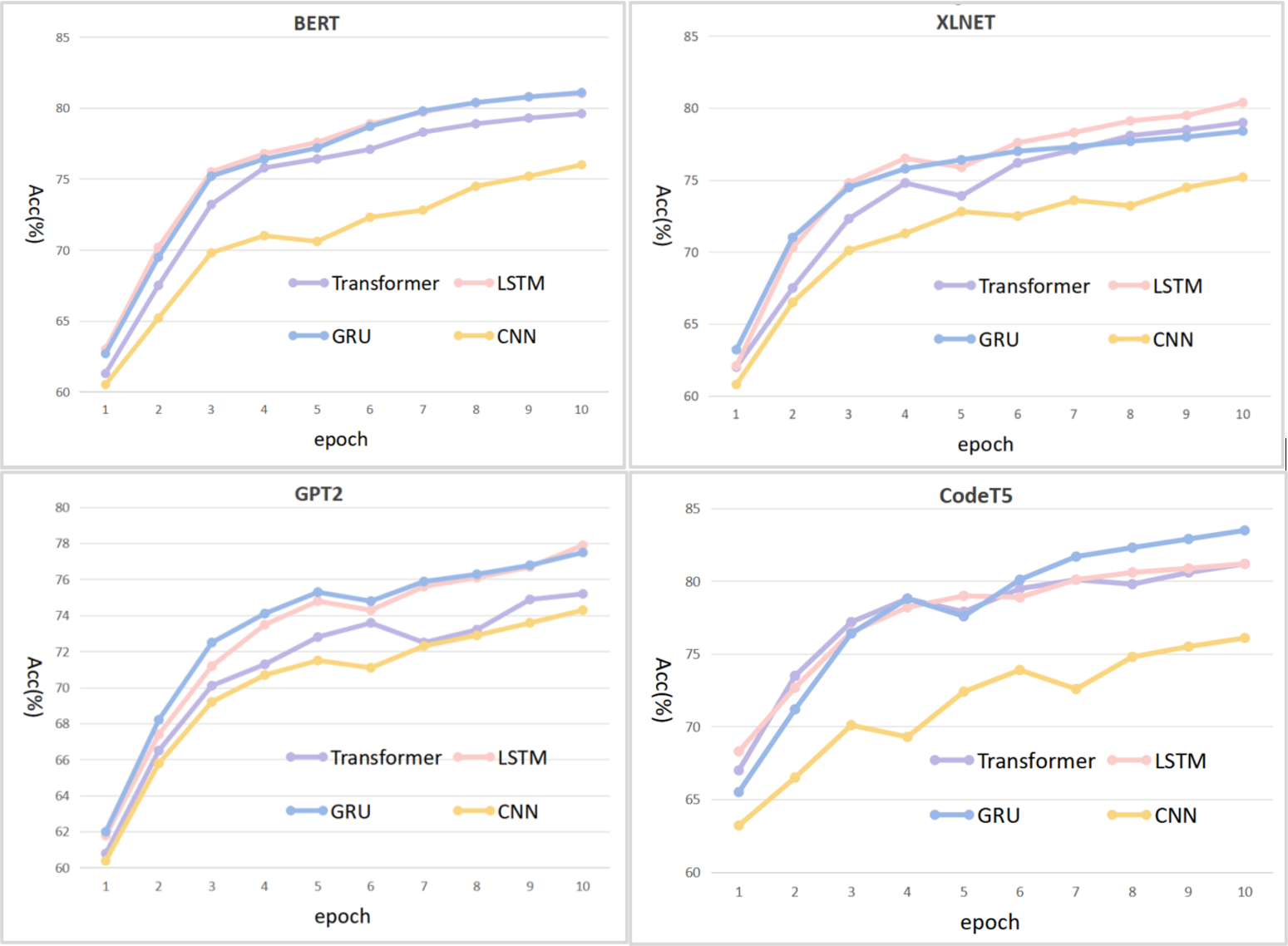}
\caption{Training Curves of SIFT-VulMiner Across BERT, XLNET, GPT2, and CodeT5} 
\label{LAT}
\end{figure*}

\begin{figure*}[htbp]
\centering
\includegraphics[width=1.0\linewidth]{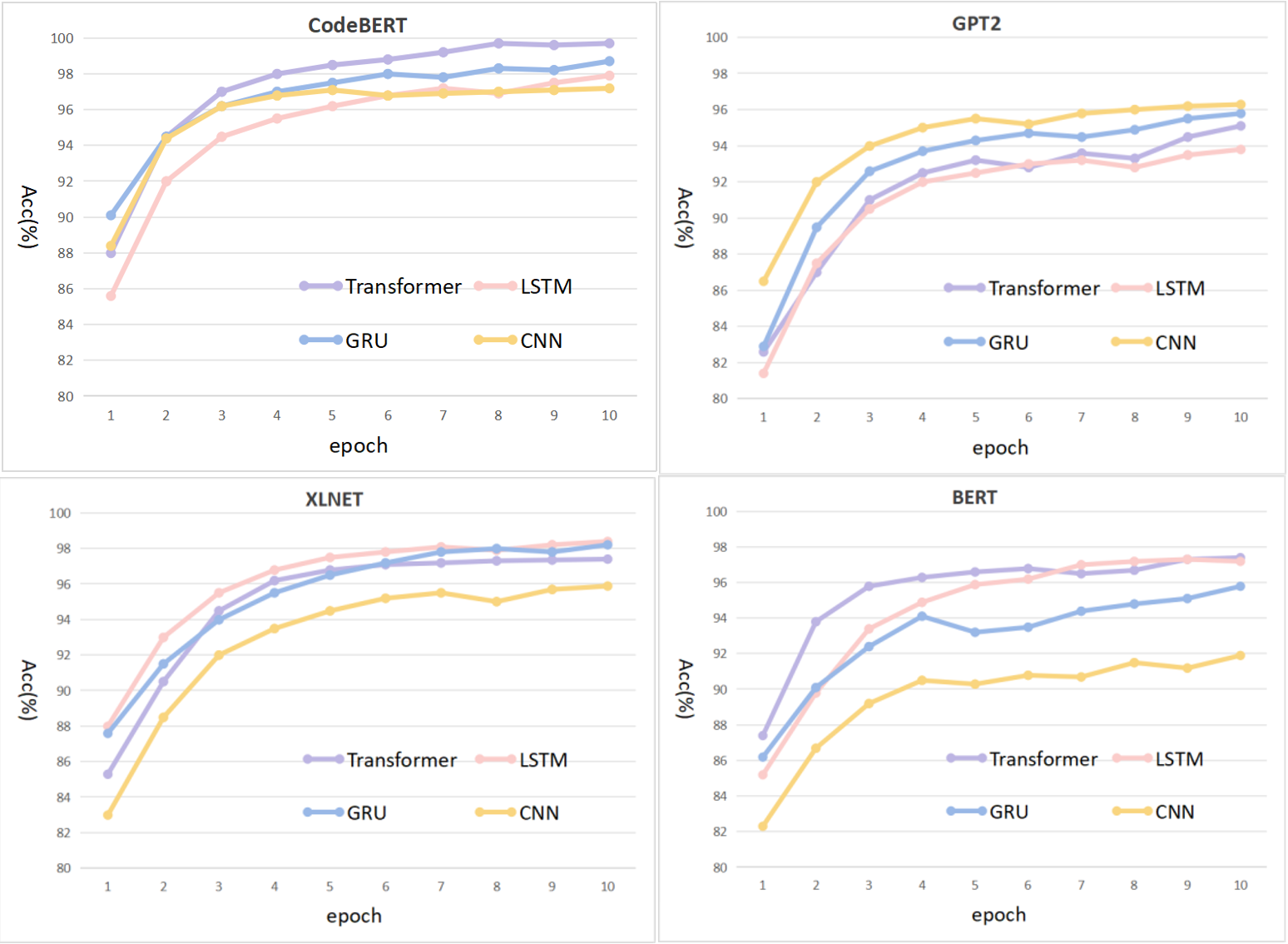}
\caption{Training Curves of SAFE-VulMiner Across BERT, XLNET, GPT2, and CodeT5} 
\label{CBT}
\end{figure*}

For the training curves of our model, please refer to the visualization results in Figure \ref{LAT} and Figure \ref{CBT}.

\section{Open Science}

We have publicly released part of the source code of AutoVulnPHP at \url{https://anonymous.4open.science/r/AutoVulnPHP-CF41/}. To curb potential misuse risks, we have implemented a controlled access mechanism, and the complete code and training datasets are only available with authorization.
Those interested in using the resources may submit an access request via an institutional email, specifying their identity, affiliation, and specific intended use. We will review all submitted requests to ensure that the relevant resources are used for legitimate academic research or industry-driven cybersecurity (and related fields) research, thereby promoting technological progress while preventing their malicious exploitation.

\section{Ethical Considerations}
All analyses and model executions were performed in isolated, non-routable environments. Discovered vulnerabilities were reported to maintainers through a responsible-disclosure process and withheld from public release until fixes were available. We do not release exploit-capable payloads or unpatched PoCs; the dataset is sanitized that forbid offensive applications.

We encourage future users to obtain authorization before scanning third-party code, to maintain audit logs, and to prioritize patching over exploitation. The goal of this work is to improve automated vulnerability discovery and repair, not to enable misuse.


\end{document}